# Mitigating biases in big mobility data: a case study of monitoring large-scale transit systems

Feilong Wang[a], Xuegang (Jeff) Ban[b], Peng Chen[c], Chenxi Liu[b], Rong Zhao[d*]

[a]School of Transportation and Logistics, Southwest Jiaotong University, Chengdu, China
[b]Department of Civil and Environmental Engineering, University of Washington, Seattle, WA, USA
[c]School of Public Affairs, University of South Florida, Tampa, FL, USA
[d]Department of Industrial Engineering, Dongguan University of Technology, Dongguan, China

**Abstract**

Big mobility datasets (BMD) have shown many advantages in studying human mobility and evaluating the performance of transportation systems. However, the quality of BMD remains poorly understood. This study evaluates biases in BMD and develops mitigation methods. Using Google and Apple mobility data as examples, this study compares them with benchmark data from governmental agencies. Spatio-temporal discrepancies between BMD and benchmark are observed and their impacts on transportation applications are investigated, emphasizing the urgent need to address these biases to prevent misguided policymaking. This study further proposes and tests a bias mitigation method. It is shown that the mitigated BMD could generate valuable insights into large-scale public transit systems across 100+ US counties, revealing regional disparities of the recovery of transit systems from the COVID-19. This study underscores the importance of caution when using BMD in transportation research and presents effective mitigation strategies that would benefit practitioners.

[*] Corresponding author: zhaorong@dgut.edu.cn

1. INTRODUCTION

Big mobility data (BMD) has been popular for monitoring the performance of transportation systems and supporting policy-makings (*1*). Here BMD refers to mobility data passively collected and generated, often by a third-party vendor (e.g., Inrix, Cuebiq, Apple, Google, and TomTom), for (often non-transportation related) primary purposes such as phone calls, shopping, entertainment, etc. (*2*, *3*). These BMD are highly attractive as they could provide a near-real-time picture of evolution in human activities across nations at no (or vastly reduced) cost for users (*4*, *5*). However, despite a number of advantages and the broad impacts generated by BMD, a fundamental question remains inadequately addressed: how BMD would be biased in terms of representing the true travel pattern and whether the data users could effectively mitigate the biases. This study centers on answering this fundamental question.

The biases are related to two limitations associated with BMD. *First*, the data properties depend on the underlying data collection methods, which are primarily non-transportation-related and largely unclear to data users (*2*, *3*). BMD only cover a sample of the population. The sample is unlikely probabilistic, because the population contained in the data is not probabilistically selected but self-opt into using specific services that contribute to the data collection. Consequently, the dataset can be biased and non-representative of the underlying population, even though it is big in the "sample size." For example, Google collects mobility data by sighting relevant locations (e.g., transit stations) of their users, while Apple collects data by counting the number of requests exclusively made to Apple Maps for directions (*6*). *Second*, in many cases, there is a lack of transparency about the methodology for data processing (*7*). The data's ability to capture mobility patterns depends on how data providers address a number of factors such as usage patterns of services that generate the data, signaling channels (e.g., cellular triangulation vs. GPS), and operational characteristics that vary with the providers (*8*). For example, how Google defines that a person visits a transit station and to which extent the definition accurately represents human mobility are not disclosed. The readers can refer to our previous reports (*3*, *9*) for more illustrations of biases in BMD.

Despite these potential issues, it is common to find studies that directly utilize BMD for human mobility and traffic activity patterns without proper validation and bias mitigation (*10*), resulting in misleading findings. This study illustrates the *biases in BMD* and focus on the recently published data reflecting transit use patterns. For monitoring the performance of transit systems, (conventional) transit data are also accessible from local transportation agencies and can serve as benchmark data. By comparing transit use patterns obtained from BMD with the data from agencies (shorten as *agency data* for convenience), we highlight cases where BMD is inconsistent with agency data and quantify the biases that may be induced using BMD for various transportation applications.

We proceed to explore *potential mitigation methods* for correcting BMD and assess the promises of BMD post-mitigation. Our findings reveal that the commonly employed bias mitigation method—linear scaling—is ineffective for BMD. The challenge arises from the varying linear relationship used for scaling across regions and its evolution over time, making it challenging to determine appropriate scaling factors for bias correction. However, through further investigation, we show that if BMD present high correlations with benchmark data, a common data processing technique—data standardization, is effective to mitigate the effects of data biases. Using transit data from Google and Apple data as a case study, we show that the mitigated BMD following data standardization uncovers valuable insights into large-scale public transit recovery across more than 100 US counties, demonstrating the effectiveness of the proposed mitigation method. By integrating county-level socio-economic, built environment, and demographic data, we identify associations between transit recovery from COVID-19 and county-level factors, some exhibiting non-linear relations. Notably, it is found that counties with higher population density and house occupancy rate recover slower, while more disadvantaged households recover faster.

BMD offers numerous advantages due to its widespread availability, including a deeper understanding of large-scale travel patterns, real-time traffic management, improved predictive modeling, and data-driven decision-making for transportation systems. While agency data could be available, its limited accessibility (e.g., one may need to communicate with each agency to access data) and update frequency restrict its usefulness in timely monitoring large-scale impacts and transportation system recovery during events like Covid-19. The comparative study of BMD and benchmark data reveals data biases and provides insights into the strengths and weaknesses of different data sources concerning data quality. The proposed mitigation methods offer valuable guidance for practitioners properly using BMD in assessing the performance of transportation systems.

In the rest of this paper, we provide a systematic literature review on potential biases in BMD and mitigation methods in Section 2. Section 3 introduces BMD and benchmark datasets used for demonstrating the proposed methods. In Section 4, we focus on quantitively evaluating the biases in BMD and show how the biases impact decision-making. The mitigation method is proposed in Section 5 and the promises of the mitigated data are illustrated in Section 6. Lastly, in Section 7, we conclude the work, and the limitations of the proposed methods and future directions are discussed. We note that this work specifically focuses on big mobility data and the application of monitoring the performance of transit systems. Codes and data in this study are available to facilitate tests on other types of data and applications (see Data Availability).

## 2. LITERATURE REVIEW

Biases in BMD and mitigation methods are reviewed, emphasizing the urgent need to address the biases to reduce the misuse. In particular, we highlight the challenges and methods related to addressing biases in BMD, which are crucial for accurate and reliable analysis in transportation and related fields.

### *2.1. Biases in Big Mobility Data*

BMD have been popular in real-world applications and policy-making (*11*, *12*). However, BMD are often passively generated via a non-probabilistic process. Consequently, BMD can be biased and non-representative of the underlying population, even though it is big in the "sample size." Empirical studies show that mobile phone users are unevenly distributed in gender and geography and population dimensions (*13*, *14*). Using the demographic information associated with mobile phone users, Yuan et al. examined how age and gender impact the correlation between mobile phone usage and mobility patterns derived (*15*). Uğurel et al. investigated biases in mobile phone data due to its sparsity in space and time and developed a Gaussian-based method for bias mitigation (*16*). Using a survey data containing information on mobile phone ownership and usage and socioeconomic status in Kenya, Wesolowski et al. showed that distinct regional, gender-related, and socioeconomic variations exist in phone usage of Kenya citizens, with particularly low ownership among rural communities, people of poor education and income (*17*). The authors analyzed how users' mobility patterns derived from mobile phone data could be correlated with socioeconomic statuses, mobile phone ownership and usage patterns of phone users. In terms of phone usage, Wesolowski et al. identified a positive relationship between income and spending on phone usage (*18*). Recently, Wang et al. found that an inadequate processing of mobile sightings data (containing call detail records data) can lead to the overestimation of a fundamental characteristic of human mobility patterns (*7*). Social media data, such as geotagged Twitter data, have also been popular for mobility analysis. Unfortunately, such data could also be biased, despite their data volume and extensive spatial and temporal coverage. Mislove and Lehmann found that Twitter users significantly over-represent the densely populated regions of the U.S., and are predominantly male (about 70%) (*19*). Meanwhile, it was found that Twitter users are biased in terms race/ethnicity in various US regions. Geographically, Malik et al. reported that more locational data are collected from Twitter in the coastal area (the east or west coast) of the US than these from other regions, which could be related to the high population density in the coastal area (*20*). Besides the Twitter data, Hecht and Stephens analyzed four-squared POI data and found these data have more users, more information, and higher quality information within metropolitan areas than non-metropolitan areas (*21*).

*2.2. Mitigations Methods*

A few recent studies attempted to address biases in BMD (e.g., social media data and web data) (*22–24*). A common procedure applied to correct biases is to apply sample weights, which are simply adjusting factors that indicate the number of units in the population that a sampled unit represents. These methods are initially proposed for addressing survey/experimental data by weighting non-probabilistic samples to correct data biases. The adjusting weights are calculated by comparing the distribution of BMD and benchmark, external data. Then, the identified biases are related with a set of independent variables; common independent variables are socio-demographics. Biases are calculated as the difference between the biased sample (e.g., mobile sensor data) and a ground truth sample. Zagheni and Weber proposed to assume and learn a model expressing biases in the estimate from a biased internet data source (e.g., web searches for a particular keyword, or geo-located tweets) (*25*). The ground truth sample is from a reliable data source (e.g., a census or a large representative survey), from which the ground-truth estimate is obtained and compared with the biased estimate. The difference is then used as the dependent variable and is related with a set of demographic variables to learn a model. The challenge associated with these weighting methods is that they require independent variables (e.g., socio-demographic features) to be available. Additionally, even they are available, it is non-trivial to identify how a set of socio-demographic variables may be related to biases, as there is a significant amount of uncertainty involved due to the complex process required (*26*).

To relieve the challenges, it is common to simplify the weight methods by identifying the relationship between BMD and the benchmark data (*27*). Then parameters describing the identified relationship can be used to correct biased estimates from a dataset that is collected from another study region or period. A simple linear relationship has been often identified and the biased BMD could be corrected via linear rescaling (*23*). However, as we show in Section 4 that the linear relationship can vary across regions and over time. This suggests that the discovered relationship in one study region or at certain time may be invalid for bias correction in another region or at another time, making the rescaling method less attractive.

Mitigating data biases is an active research area in machine learning community, where data bias refers to the problem of certain population groups in training data not being representative of the actual data population. Consequently, the trained model could exhibit bias towards certain groups (e.g., minorities) during the inference stage, as measured by fairness metrics. The computation of these fairness metrics requires labels or population attributes (e.g., patients' personal information in (*28*)), which are typically absent in mobility data. As noted in (*29*), current methods tend to rely on high-granularity data attributes, which would limit the applicability of these methods. BMD data presents unique challenges due to the unknown data generation mechanism, making these machine learning methods not directly applicable. Specifically, data in the machine learning field often contain ground-truth labels and attributes that can be associated with the data generation process for correction purposes (*30*).

## 3. DATA DESCRIPTION

**Google Data** We collect and process two widely used open-sourced BMD, including Apple mobility reports and Google community mobility reports. Google mobility reports use aggregated, anonymized data to derive movement trends across six different categories in "transit stations", "retail and recreation", "groceries and pharmacies", "parks", "workplaces", and "residential" (*6*). The data are collected as people are sighted staying at relevant locations such as transit stations. As such, it does not directly inform us about transit boardings or ridership. Google defines the baseline of the dataset as the median value for the corresponding day of the week, during the 5-week period from January 3 to February 6, 2020. Then, relative daily changes from the baseline at the county level are provided. The data is updated twice a week. Here, we focus on data related to "transit stations".

**Apple Data** The Apple mobility reports cover major regions in the globe. Different from Google who collects data by sighting users' locations, Apple data are generated by counting the number of requests made to Apple Maps for directions (*31*). Despite the difference in measurement, it has been shown that the daily changes in navigation requests (e.g., from Apple Maps) also offer useful insights into people's mobility changes in response to COVID-19. Apple provides mobility records in three categories that include "transit", "walking", and "driving" and this study focuses on their "transit" data. In addition, Apple defines mobility from January 13 as the baseline value. Similarly, relative daily changes from the baseline at the county level are derived. Neither Google nor Apple collects user demographic information. Apple Maps is only available on Apple devices. Therefore, it is unknown whether the users included in the data are representative of the entire population.

**Transit data from local agencies (AD)** Local transit agencies actively collect ridership information for each transit route. We focus on the boardings data they resemble the closest to the Google and Apple data. Note that AD is an estimate of transit ridership based on counters installed on a sample of transit vehicles and thus would contain errors. AD is relatively trustworthy and treated as the *benchmark data* (*32*).

**Data Collection and Preprocessing** For comparing BMD and AD, we first compare them in four regions in the US, including King County (KC) in Washington, Los Angeles area (LA) in California, New York City (NYC) in New York, and Washington Metropolitan area (WM) in the District of Columbia, as the case studies to help assess the quality of BMD discussed above. These four regions are chosen since they help cover different regions of the US (especially the east coast and the west coast) and it is relatively easy to access the transit data in the four regions. Diverse seasonal climate behaviors are also covered, for example, the strong and weak temperature seasonality in NY and in LA, which could play a role in the performance of transit systems. The bias mitigation method will be proposed and applied in all the US counties (138 counties in total) having transit data from BMD.

The study period ranges from January 2020 to April 2022, when Apple stopped providing data updates. Several preprocessing steps are necessary to make the datasets more comparable, including imputing the missing values on two days, aligning the baselines and aggregating daily data into monthly for comparisons. Another dataset from Transit (a popular mobile phone app for real-time public transit data and trip planning) is also collected and analyzed (*46*). Readers are referred to Appendix B for the data description and results from testing the proposed methods.

## 4. DIFFERENCES BETWEEN BMD AND AD AND THEIR IMPACTS

We begin with quantifying the disparities in *values* and then evaluate whether the two types of data exhibit similar *trends* patterns, admitting the fact that BMD possessing similar trends as AD could still provide valuable insights despite the disparities in values. We also show how disparities impact decision-making.

### 4.1. Difference between BMD and AD

Fig. 1 shows the trends of the transit usage in the four regions derived from BMD and AD. Following the *difference-in-difference* technique, all the comparisons rely on the percent changes from the (pre-covid) baseline. For example, -80% here means that the transit usage dropped by 80 percent compared with the baseline. The comparison reveals some *similar* patterns between AD and BMD. Specifically, all datasets show an abrupt drop in the first quarter of 2020 when some restrictive mobility-related regulations took place as COVID-19 started in the US. All datasets witness a gradual increase in transit use during the past two years with several ups and downs, which are likely driven by several waves of COVID-19 variants. In the following, the differences between the two types of data are qualified and analyzed.

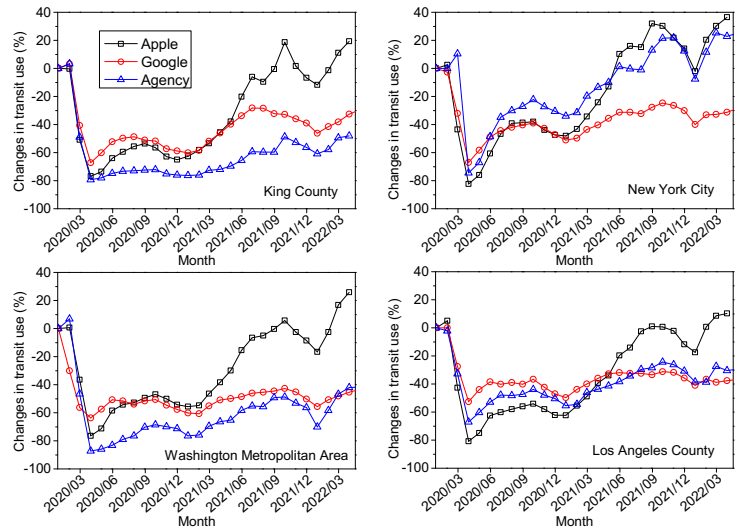

**Fig. 1**. A general comparison of BMD and AD.

Table 1 presents the average of the values (changes from pre-covid baseline) calculated from BMD and AD, with standard deviations as error bars. The average values suggest the average impacts of COVID-19 in terms of reducing transit usage during the study period (e.g., a -46% by Google data at KC means an average reduction of 46%). The deviations tend to be spatially heterogeneous. On average, Google data underestimate the impact by 20% at KC, 14% at WM, 3% at LA and overestimate the impact by 26% at NYC. Apple data overestimate the impact by 33% at KC, 36% at WM, 8% at LA and underestimate the impact by 2% at NYC.

Table 1. Mean and standard deviations of BMD and AD for monitoring changes from pre-covid baseline.

| Region | Agency | | Google | | Apple | |
| --- | --- | --- | --- | --- | --- | --- |
| | Mean | Standard deviation | Mean (under-/over-estimation) | Standard deviation | Mean (under-/over-estimation) | Standard deviation |
| KC | -66 | 10 | -46 (20) | 11 | -33 (33) | 31 |
| NYC | -14 | 27 | -40 (-26) | 10 | -16 (-2) | 36 |
| WM | -66 | 12 | -52 (14) | 5 | -30 (36) | 28 |
| LA | -42 | 11 | -39 (3) | 6 | -34 (8) | 29 |

The real-time values of BMD are also useful for applications such as monitoring the real-time impacts of COVID-19 and tracking the recovery process of transportation systems. Fig. 2 shows the real-time (instant) deviation of BMD from AD in the four regions monthly. Note that for both types of data, their percentage changes from the pre-covid baseline (see Fig. 1) are used for computing the instant deviation. A positive (negative) deviation at time $t$ indicates that BMD estimates more (less) transit usage than AD does and thus underestimates (overestimates) the impacts. Surprisingly, the instant deviation of BMD not only varies across the regions but also depends on the time period of investigation: the deviations in all the regions demonstrate noticeable (either increasing or decreasing) trends instead of random fluctuations around a constant bias (i.e., the average deviation). The deviation in Apple data starts growing around March 2021. On the other hand, the instant deviation in Google data tends to drop over time. For Google data, the underestimation (overestimation) ranges from -60 to 30; for Apple data, it ranges from -20 to 60.

These findings suggest that BMD provided by different vendors may be consistent with or deviate from AD more, depending on the specific regions and specific periods of time. For example, it appears that Google data agrees more with AD in WM area especially after 09/2021, while Apple data is more consistent with AD in NYC. In applications, BMD that is consistent with AD at one location at a specific time may not imply the same in another region or at another time period. Since the methods of collecting data and estimating transit use are largely unknown from BMD providers, it is difficult to investigate the reason for these unexpected patterns, raising caution for researchers and practitioners while using BMD solely for real-time applications and decisions.

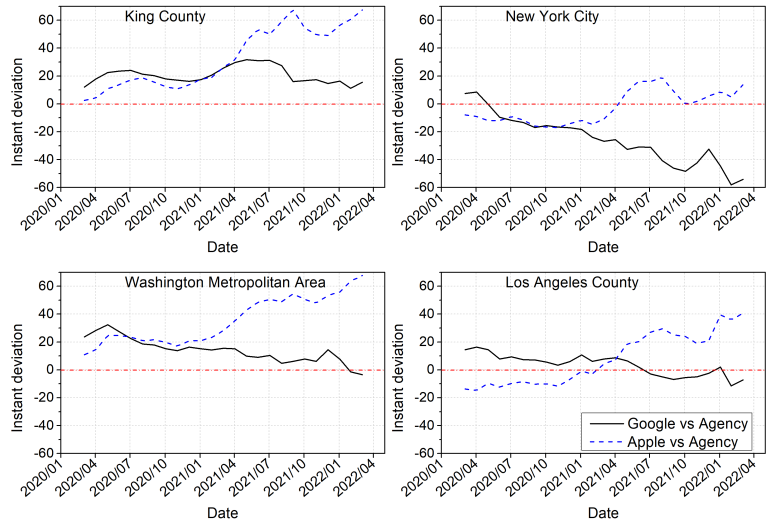

**Fig. 2**. Monthly deviations of BMD from AD.

*4.2. Impact of Differences in Data on Decision Making*

The analyses conducted thus far reveal some similarities and many differences between BMD and AD. This raises the question of how these differences impact decision-making when utilizing BMD. Previous studies have often relied on BMD as a direct measure to estimate the impact of COVID-19 on mobility. However, a closer examination of Figures 1-2 and Table 1 reveals substantial differences between BMD and AD. Using BMD directly for transportation assessment would lead to either overestimation or underestimation of the impact. Fig. 2 highlights the presence of temporal and spatial heterogeneities. It turns out that neither Google nor Apple data is consistently closer to AD across all the regions and all the time periods. Consequently, if BMD is employed to forecast the spread of COVID-19 in communities, the reliability of such forecasts becomes questionable unless the deviations in BMD are properly acknowledged and addressed.

We then evaluate the value of changing trends and patterns inherent in BMD for decision-making purposes. One common application of BMD is to assess the impact of a COVID-19 variant or the effectiveness of safety-related policy interventions using real-time information derived from this data. For instance, transit agencies may use BMD to evaluate whether social distancing or face-masking regulations are discouraging ridership by examining the presence of a decreasing trend. In such cases, the trend itself, rather than the absolute values, can provide valuable insights even if the absolute values may not be reliable. This type of analysis empowers transportation agencies to anticipate potential shifts in demand, enabling them to adapt policies and allocate necessary resources to ensure system readiness. However, our analysis below suggests that potential errors may arise when utilizing the trends in BMD for assessing the impacts of policies or COVID-19 waves on transit systems.

The key step here involves detecting change points in the time series, which signify significant changes in the underlying properties of the data, often resulting from important events or transitions like policy changes or natural disasters. The output of change point detection is typically a set of change points along with some statistics (e.g., the likelihood of the detected changes). These change points can be further utilized for analysis, or decision-making purposes, enabling one to understand and respond to the identified shifts in the time series data. Many recent applications related to COVID-19 have relied on this approach, as exemplified in (*33*), to identify and evaluate the time lag effects of COVID-19 policies on human mobility. This study applies the widely used Bayesian change point detection method (BCPD) to analyze both BMD and AD and compare the results. BCPD computes the posterior probability of a change point occurrence at a specific time point; details are omitted due to space limit.

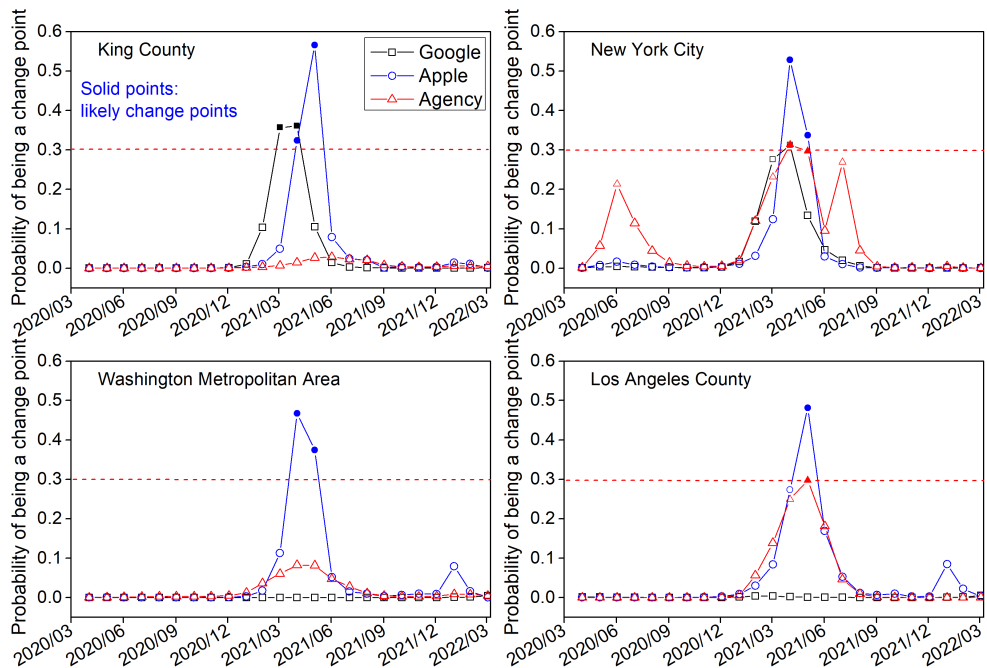

**Fig. 3**. Change point detection to support decision-making using original data.

In Fig. 3, the curves represent the probability of being a change point, while the solid symbols indicate the identified change points using threshold of 0.3. It can be observed that neither Google nor Apple data is consistent with AD data in terms of change point detection results. More specifically, from the AD data, we observe change points detected around the spring of 2021 in LA and NYC, while no change points are detected in KC or WM due to their relatively steady and gradual recovery (Fig. 1). However, Google data completely misses the change points in LA and WM. On the other hand, Apple data tends to produce false-positive errors, although it successfully detects change points in LA and NYC. It is worth noting that such false-positive errors are common in search engine-based datasets (including the Apple data here), which are susceptible to fluctuations in search activities (*34*).

## 5. MITIGATION METHOD

Section 4 highlights that despite employing the difference-in-difference technique, BMD can exhibit substantial differences compared to AD and the differences affect decision making. Here we investigate potential bias mitigation methods. Mitigating biases in BMD has important implications. Though agencies can use AD data for performance evaluation (if available), AD data has limitations that could be complemented by BMD that is spatially ubiquitous with frequent updates. This means that BMD tends to be available for evaluating transit performance at places where AD data are lacking. Therefore, a mitigation method for BMD is valuable to complement AD data.

### 5.1. Failure of Linear Rescaling as a Mitigation Method

We first test the *linear rescaling method*, a commonly used mitigation method for BMD. It is to rescale BMD following two steps.

1. It fits a linear function between BMD and AD:

$$AD(t) = \beta_0 + \beta_1 \cdot BMD(t) + \varepsilon(t) \quad (1)$$

2. The fitted parameters (i.e., intercept $\beta_0$ and slope $\beta_1$) are applied to rescale BMD:

$$ScaledBMD(t) = \beta_0 + \beta_1 \cdot BMD(t) + \varepsilon'(t) \quad (2)$$

Here, $\varepsilon(t)$ and $\varepsilon'(t)$ are the residuals of fitting and rescaling BMD at time $t$, respectively. Despite the simple steps, we show below that implementing the rescaling method in practice may not be trivial.

The scatter plot in Fig. 4 shows the correlation between BMD and AD, where each dot represents the monthly change of transit usages from the pre-covid baseline (see Fig. 1). We employ a color scheme to account for the time-dependent nature of the differences between BMD and AD (Fig. 2): the dots are darker the further they are from the pandemic break (Feb. 2020). We first measure their Pearson correlation (*35*). Remarkably, Apple data exhibit a consistently strong correlation with AD, with correlation coefficients exceeding 0.94 across all regions. On the other hand, Google data show less consistency with AD, as indicated by the scattered distribution of dots and lower correlation coefficients. Meanwhile, the relationship between BMD and AD appears to be unique to each region. This spatial heterogeneity raises concerns about estimating the *global* impact of COVID-19 on transportation systems based on a relationship derived from a *single* region (*36*). As shown latter, the varying relationships also make it challenging to estimate the impacts of COVID-19 using linearly rescaled BMD (*37*, *38*).

The region-dependent relationship also means that the relationship estimated in one region is not generalizable to other regions (or even other time periods of the same region), suggesting that existing studies using learned relationship from specific region for the global estimate would likely be problematic. Moreover, our previous investigation found that the relationship could evolve over time. Specifically, fitting the data at different time points of investigation, the fitted parameters evolve over time. The temporal and spatial variations of the linear relationship mean that the parameters for scaling BMD need to be corrected using the up-to-date and location-specific AD. However, the heavy reliance on AD renders the linear scaling less helpful as a mitigation method.

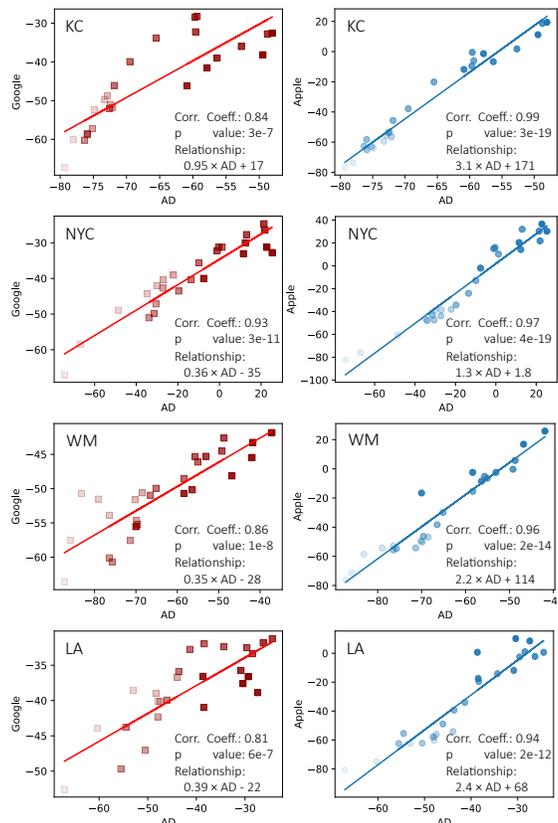

**Fig. 4**. Correlation relationship between BMD and AD.

*5.2. Proposed Mitigation Method*

The analysis above highlights that despite significant variations in values between BMD and AD, both data types demonstrate linear relationships, indicating BMD's potential to capture the overall trend of transit usage. Leveraging these linear relationships, we can develop a mitigation method. Through standardization, a common data processing technique, BMD can be converted to align with AD. Standardization offers multiple advantages, including data normalization while preserving data information.

Given the linear relationship $x^i_{BMD} = ax^i_{AD} + b$ (Fig. 4), we have standardized data $\hat{x}^i_{BMD} = \frac{x^i_{BMD} - \mu_{BMD}}{\sigma_{BMD}}$ and $\hat{x}^i_{AD} = \frac{x^i_{AD} - \mu_{AD}}{\sigma_{AD}}$, which are equivalent (i.e., $\hat{x}^i_{BMD} = \hat{x}^i_{AD}$). See proof in the appendix. Here, $\mu$ and $\sigma$ represent the mean and standard deviation of BMD and AD, respectively. This property ensures that $\hat{x}_{BMD}$ that is independently derived without relying on agency data preserves the same information as $\hat{x}_{AD}$. Such independence alleviates concerns associated with the linear scaling method. Fig. 5 demonstrates the consistency of $\hat{x}_{BMD}$ with $\hat{x}_{AD}$. Notably, the standardized Apple data closely aligns with the standardized AD. Since Apple and Google have different data collection processes and properties, this observation suggests the generalizability of data standardization as an effective mitigation method (see limitations discussed in Section 7). Given our focus on the promises of the trend patterns, we skip quantifying the instant deviations between $\hat{x}_{BMD}$ with $\hat{x}_{AD}$.

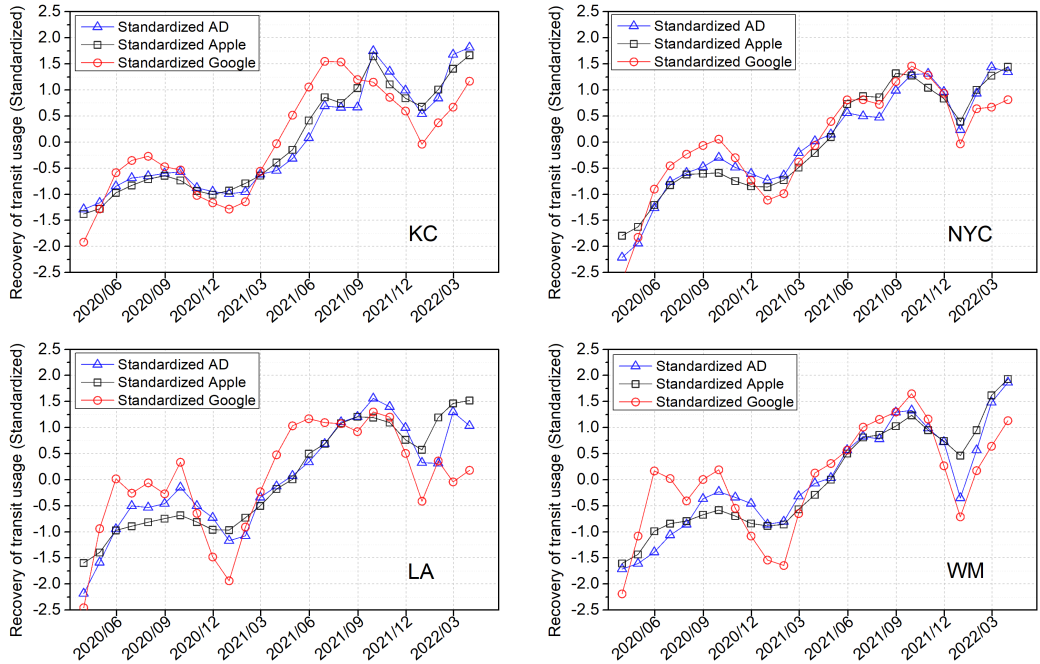

**Fig. 5**. Comparison of the standardized BMD and AD

6. PROMISES OF MITIGATED DATA

We utilize the Bayesian change point detection method once more on the mitigated BMD and AD, and the results are depicted in Fig. 6. Compared with the outcomes without data mitigation, the change point detection results from mitigated BMD and AD exhibit greater consistency, indicating the effectiveness of standardization as a data mitigation technique. We note that the likelihood associated with standardized AD is at the different scale with that from the original AD due to the standardization and updated hyperparameter of the change point detection algorithm. Moving forward, we proceed to demonstrate the value of mitigated BMD in examining the recovery process of transit systems across the US and identifying potential associations with demographic and socio-economic factors. Note that since Apple data after mitigated is more in line with AD, our demonstration is focused on Apple data.

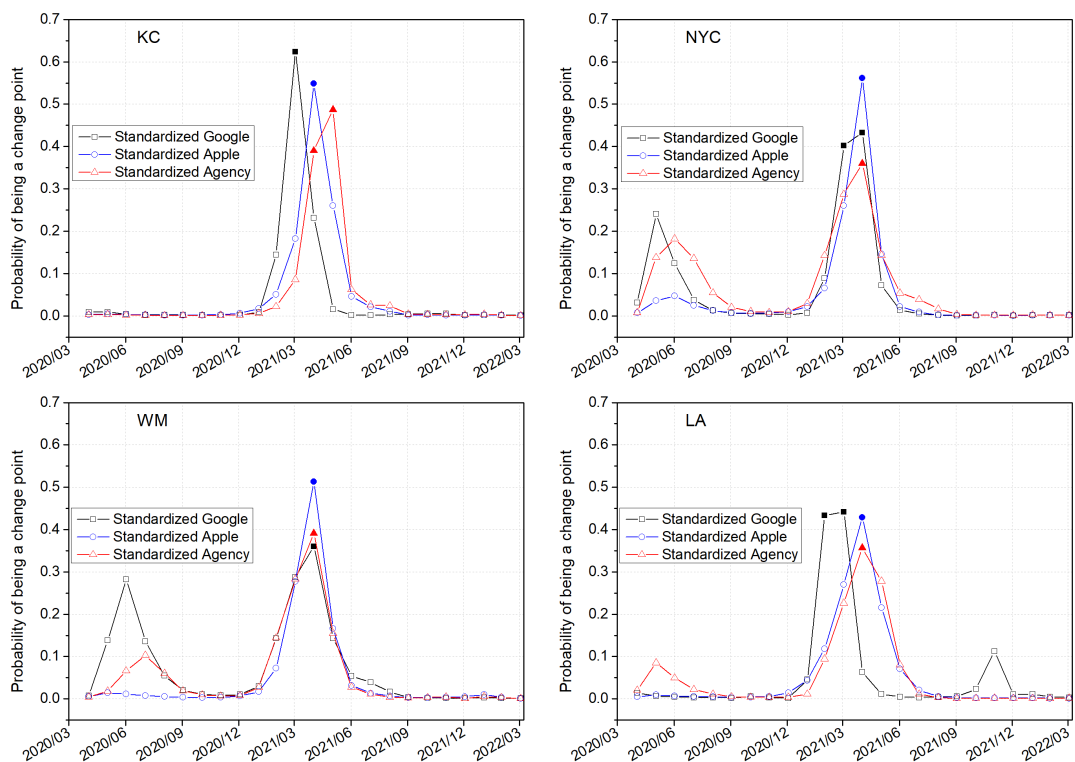

**Fig. 6**. Change point detection to support decision-making using standardized data

With the mitigated data, we can delve into the recovery process of large-scale transit systems in the United States. To achieve this, we employ a Logistic growth curve (Equation 3) to fit the observations within each county. The *fitted growth rate* ($k$) serves as an indicator of the transit systems' resilience during Covid-19. A higher value of $k$ corresponds to a faster growth rate, indicating a more resilient transit system.

$$ScaledBMD(t) = \frac{L}{1+e^{-k(t-t_0)}} \qquad (3)$$

Since $k$ does not change with the data standardization, $k$ learned from $\hat{x}_{BMD}$ aligns with that from $x_{AD}$, which is challenging to collect for all US counties. This study focuses on the recovery process in 2021, as riders gradually returned to work, and transit system schedules recovered from disruptions. The example in Fig. 7A demonstrates a well-fitted Logistic growth curve for KC in 2021, yielding a $k$ of 0.95. Fig. 7B illustrates a bell-shaped distribution of recovery rates across US counties, with a mean of 1.3 and a standard deviation of 1.6. Notably, recovery rates vary among counties; for instance, Bronx County, NY, and Queens County, NY exhibit a recovery rate of 1.2, while San Francisco County, CA, and Los Angeles County, CA show a recovery rate of 0.8.

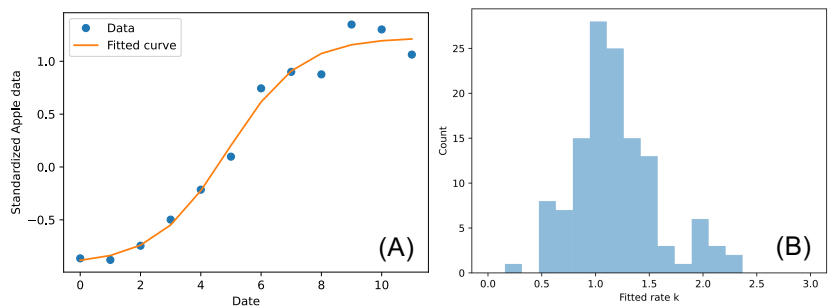

**Fig. 7**. (A) Logistic growth curve fits KC data in 2021; (B) recovery rate distribution for US counties.

To explain these variations in $k$ across counties, we employ a *Gradient Boosting Decision Trees* (GBDT) model (*39*), demonstrating that the variations are associated with multiple factors, including county-level land use data, and demographic and socio-economic features collected from American Community Survey and 2020 Census data. GBDT is popular for its several advantages. Firstly, it can effectively handle various data types, including continuous and categorical variables. Secondly, GBDT is more robust to outliers and missing data than conventional linear models. Thirdly, it addresses the issue of multicollinearity effectively (*39*). Lastly, with decision trees learned from data, GBDT can capture irregular (e.g., non-linear) associations between variables. Note that the primary objective of this study is not to identify all influential factors but to demonstrate the potential of mitigated MBD.

GBDT automatically identifies important predictors based on factor selection. Fig. 8 presents the *relative factor importance* in predicting transit recovery rates $k$, with only the top influential factors (explaining more than 80%) demonstrated. From Fig. 8, it is evident that population density alone accounts for approximately 18% of the variations in transit recovery. Given its significance in public transportation planning and operation, the role of population density in transit recovery during disruptions is expected. Other essential factors include housing characteristics, such as the percentage of occupied houses and family households with children under 18. Additionally, factors related to built environment (e.g., percentage of rural lands) play notable roles in explaining variations. While race (e.g., percentage of Hispanic population (HISP) and Native Hawaiian and Pacific Islander (NHPI)) also contributes to the variation, its impact is not as significant.

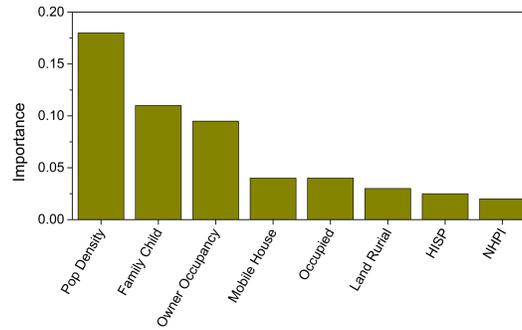

**Fig. 8**. Feature importance in terms of explaining variation of recovery rate.

GBDT also facilitates creating *partial dependence plots* (PDP) to visualize the association between influential factors and recovery rates while controlling for other factors in the model. The PDPs depicted in Fig. 9 reveal that many factors exhibit nonlinear relationships with recovery rates, helping identify their effective influence ranges. Notably, population density displays a negative association with the recovery rate. This could be attributed to social distancing practices and riders' avoidance of crowded areas, more common in high-density regions, resulting in slower recovery. This finding aligns with the previous finding (*41*) that high-density areas experienced higher impacts during Covid-19. Conversely, counties with higher house owner occupancy rates likely exhibit higher recovery rates, indicating stable and well-established residential areas with residents investing in their properties and neighborhoods. Similarly, counties with more family households with children under 18 demonstrate a likely higher recovery rate. Regarding the built environment, an association is observed from factors such as percentage of rural areas. However, the non-linearity makes it challenging to clearly interpret the effects. Additionally, counties with population in the minority group, including the percentage of HISP and NHPI, are likely associated with a slower recovery rate, suggesting the potential resilience concerns in the less represented population.

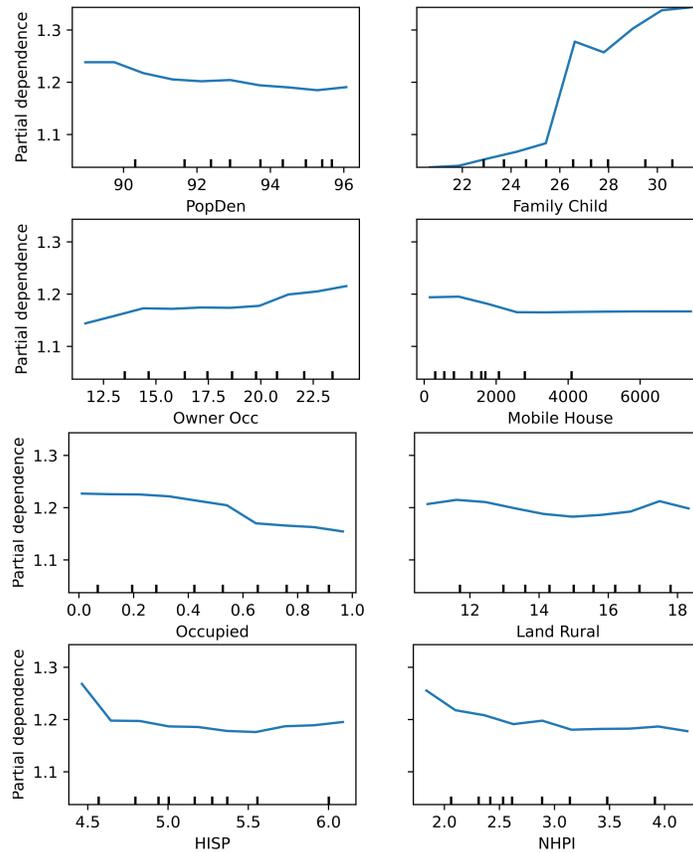
**Fig. 9**. Nonlinear relationship between demographic variables and transit recovery rate.

## 7. CONCLUSION AND FUTURE WORK

We evaluate biases in BMD using Google and Apple data as examples for monitoring transit use patterns during COVID-19. An effective bias mitigation method is then proposed and tested. By comparing BMD with AD, we identified and quantified similarities and many important differences between the two types of data, despite the difference-in-difference employed. We found that BMD can significantly overestimate or underestimate the impacts of the pandemic on transit usage. And such biases vary with the data source (either Google or Apple) as well as the study region and period. It was also observed that in terms of deviations in *values*, neither Google nor Apple data outperformed the other in all regions. The biases in the data also surprisingly evolve over time, meaning distinct conclusions may be yielded at different time points of investigation. The *trend* patterns in Apple data in general were more in line with those in AD, making it a more promising data source for trend-based analysis. The potential impacts of the biases in BMD were then investigated. It was found that besides the over- and under-estimation of the impacts of COVID-19 on the transit systems, the biases in BMD lead to false-positive and false-negative errors in the inferred change points, which are often used for making and evaluating decisions.

Potential mitigation methods are then investigated. We show that the linear rescaling as a common bias mitigation method is not effective. The linear rescaling method relies on a common linear relationship across regions, which, unfortunately, is invalid for the BMD investigated. Specifically, the linear relationship varies across regions and evolves over time, meaning no generalizable relationship is available. This renders the rescaling method unsuitable, as it needs to be site-specific and constantly updated to mitigate the biases in BMD. To relieve the concern, we utilize the high correlations between BMD and AD and prove that data standardization can be effective in mitigating biases. In the case study, we show that the mitigated Apple data uncovers valuable insights into large-scale public transit recovery across more than 100 US counties, demonstrating the effectiveness of the proposed mitigation method. Together with socio-economic, built environment, and demographic data, associations between transit recovery from COVID-19 and county-level factors can be identified and discussed.

There are several limitations with this study, demanding future research needs. First, this study focuses on examining the impacts of data biases on evaluating transit systems. More examinations of the impacts on other applications using BMD (e.g., prediction of COVID-19 cases) would benefit future practices of proper data use. Second, only Google and Apple data are examined in this study despite the diverse BMD in the field. In the future, additional types of BMD such as those from other data sources (e.g., data from smart transit card (*42*), online ride hailing (*43*) and bike-sharing (*44*)) may also be evaluated to gain a more complete picture of the potentials and biases of these BMD datasets. Moreover, it is crucial to accumulate insights and strategies for using BMD appropriately and accurately for specific applications. The findings in this paper indicate that caution should be exercised when doing so, and it is often a good practice to use alternative (and more traditional) data sources as a benchmark to evaluate and validate BMD and the results derived from BMD (*3*, *45*). Third, the validation of the proposed mitigation is limited to Google and Apple data, which exhibit linear correlations with the benchmark data. Future studies should validate the generality of the mitigation method to other types of BMD. Without tests on other types of potentially biased big data and application scenarios, determining whether the proposed method could be applied to universal big datasets and applications needs future research efforts. If BMD exhibits weak or nonlinear correlations, or other complex trends with the benchmark data, exploring the potential of machine-learning methods to model these trends could be beneficial. A challenge lies in properly validating the rescaled BMD against ground-truth data afterward. Additionally, the potential of BMD mitigated via data standardization could be explored in applications beyond change point detection and estimating recovery rate.

**Acknowledgement**

The authors acknowledge the efforts of a former student Ce Wang at University of Washington and data from collaborators at King County Metro and Sound Transit. This work was supported in part by the Guangdong Basic and Applied Basic Research Foundation under Grant 2022A1515110924.

**Disclosure Statement**

The authors report there are no competing interests to declare.

**Data Availability**

Big data and codes used in the study are available at https://github.com/activeconclusion/covid19_mobility and https://github.com/feilongwang92/mitigate_data_bias, respectively.

**Appendix**

**Appendix A. Proof of the standardized BMD and AD being equivalent.**

We show that the standardized BMD $\hat{x}_{BMD}^i = \frac{x_{BMD}^i - \mu_{BMD}}{\sigma_{BMD}}$ and AD $\hat{x}_{AD}^i = \frac{x_{AD}^i - \mu_{AD}}{\sigma_{AD}}$ (Section 5.2) are equivalent if there exists a linear relationship between BMD and AD (i.e., $x_{BMD}^i = a x_{AD}^i + b$). Here, $\mu$ and $\sigma$ are the mean and standard deviation.

Given $\mu_{BMD} = \frac{\sum_i (x_{BMD}^i)}{n}$, $\sigma_{BMD} = \sqrt{\frac{\sum_i (x_{BMD}^i - \mu_{BMD})^2}{n}}$ and assuming $x_{BMD}^i = a x_{AD}^i + b$, we have

$$\mu_{BMD} = \frac{\sum_i(ax^i_{AD}+b)}{n} = \frac{a\sum_i(x^i_{AD})+b}{n} = a\mu_{AD} + b.$$

$$\sigma_{BMD} = \frac{\sqrt{\sum_i(ax^i_{AD}+b-a\mu_{AD}-b)^2}}{n} = \frac{a\sqrt{\sum_i(x^i_{AD}-\mu_{AD})^2}}{n} = a\sigma_{AD}.$$

Inserting the two equations into $\hat{x}^i_{BMD}$, we have

$$\hat{x}^i_{BMD} = \frac{x^i_{BMD}-\mu_{BMD}}{\sigma_{BMD}} = \frac{ax^i_{AD}+b-(a\mu_{AD}+b)}{a\sigma_{AD}} = \frac{x^i_{AD}-\mu_{AD}}{\sigma_{AD}} = \hat{x}^i_{AD}.$$

**Appendix B. Test on the Transit App data.**

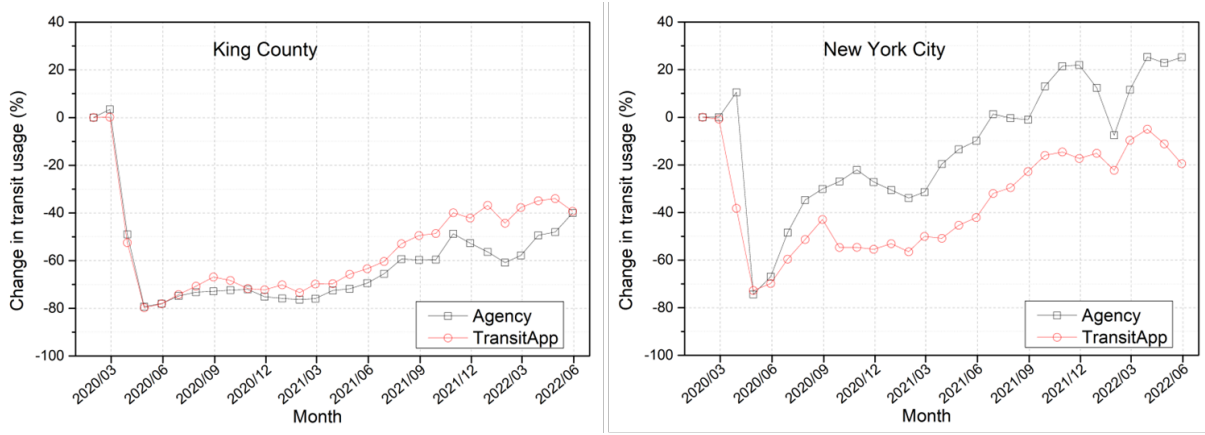

**Fig. A1**. A general comparison of Transit App data and AD.

Fig. A1 illustrates the trends in transit usage as captured by Transit App data and AD. While the comparison indicates certain parallel trends between the two datasets, it is important to recognize that the Transit App data would not accurately reflect the full impact of COVID-19 on transit performance. This potential discrepancy could result in either an overestimation or underestimation of the pandemic's effects.

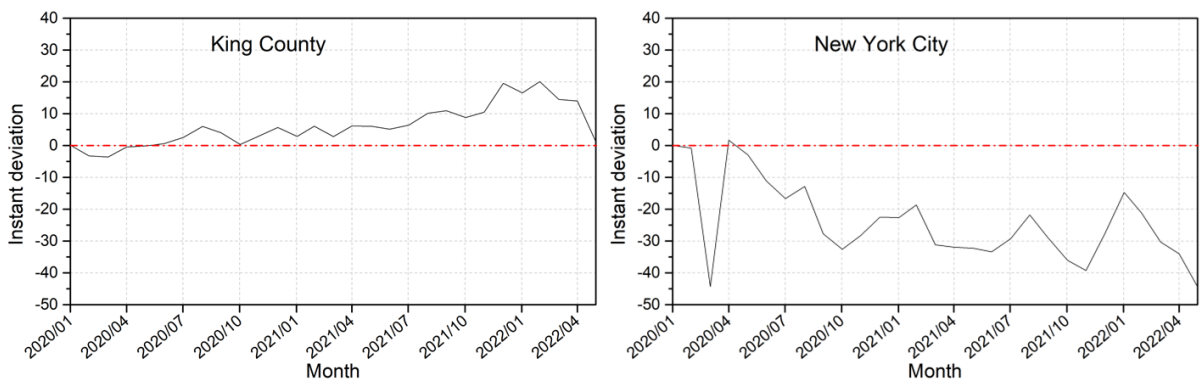

**Fig. A2**. Monthly deviations of Transit App data from AD before mitigation.

Fig. A2 displays the monthly real-time deviations between Transit App data and AD. Significant discrepancies are evident, mirroring the variations observed with Google and Apple data as discussed in the main text. These deviations in Transit App data are not uniform; they fluctuate across different regions and also depend on the specific time period under review.

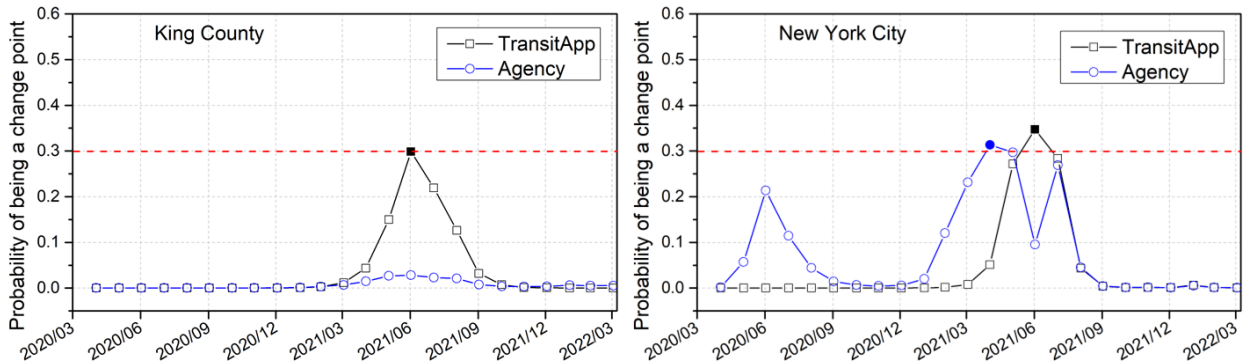

**Fig. A3**. Change point detection to support decision-making using original Transit App data. Solid points are likely change points.

Fig. A3 shows the probability of being a change point. It can be observed that without mitigation Transit App data is not consistent with AD data in terms of change point detection results, which would likely mislead policy-making.

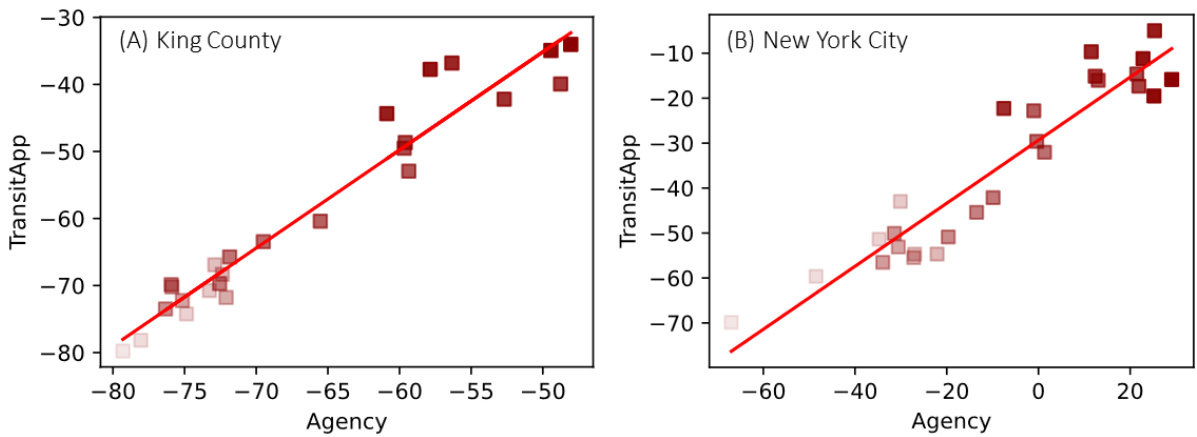

**Fig. A4**. Correlation relationship between Transit App data and AD.

The scatter plot presented in Fig. A4 illustrates the correlation between data from the Transit App and AD. The analysis reveals strong correlations between these two datasets, with correlation coefficients reaching 0.96 in King County and 0.91 in New York City.

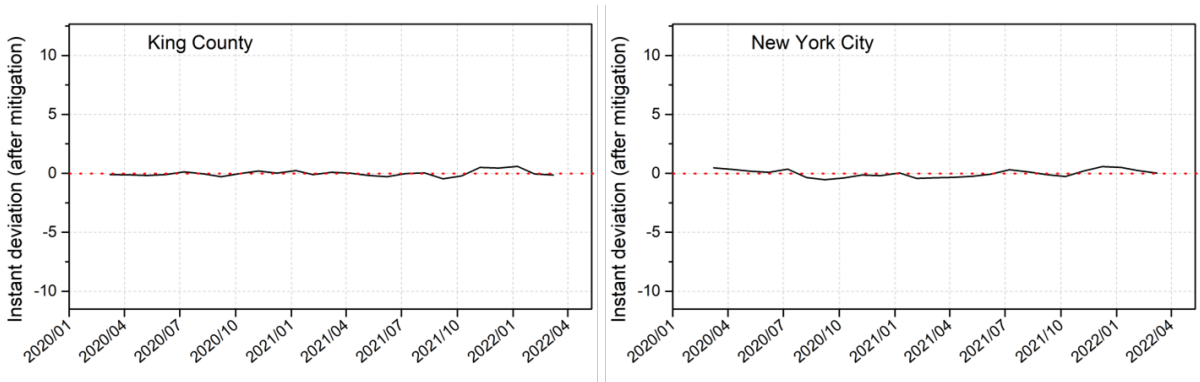

**Fig. A5**. Monthly deviations of Transit App data from AD after mitigation.

Fig. A5 shows the instant deviation of Transit App data from AD following the standardization-based mitigation. We can observe that the deviations are significantly reduced.

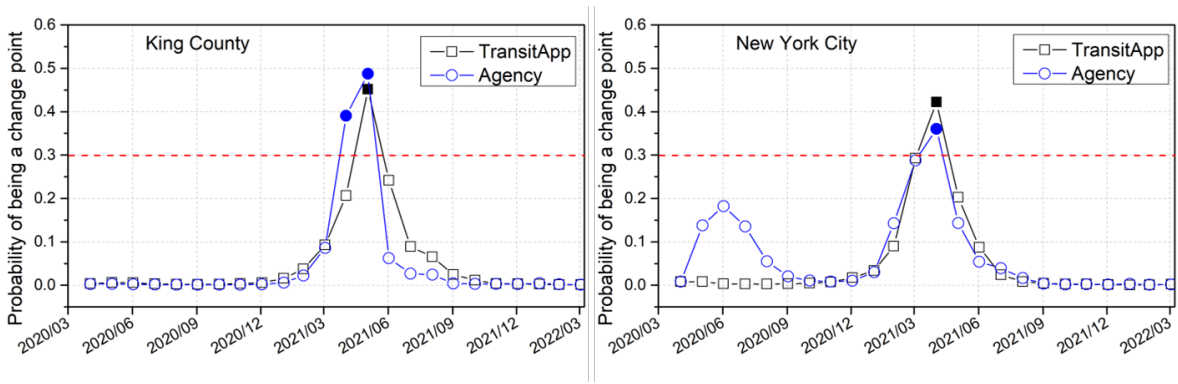

**Fig. A6**. Change point detection to support decision-making using standardized Transit App data.

Fig. A6 presents the outcomes of change point detection analysis conducted on the mitigated Transit App data and AD. In line with the findings from the mitigated Google and Apple data, the results after applying mitigation measures show improved consistency in the detection of change points. This enhanced alignment underscores the efficacy of standardization as a data mitigation technique.